# Estimating excess mortality during the Covid-19 pandemic in Aotearoa New Zealand: Addendum

Michael J. Plank[1], Pubudu Senanayake[1,2], Richard Lyon[3]

1. School of Mathematics and Statistics, University of Canterbury, Christchurch, New Zealand.
2. Statistics New Zealand, Christchurch, New Zealand.
3. Mortality Working Group, Institute of Actuaries of Australia, Sydney, Australia.

## Abstract

In our previous article, we estimated excess mortality during in Aotearoa New Zealand for 2020 to 2023. Since our work was published, updated population estimates have been released by Statistics NZ. In this short letter, we provide the results of applying our original model to the new population data. Our updated excess mortality estimate of 2.0% (95% CI [0.5%, 3.3%]) is 1.3 percentage points higher than our original estimate because the new population estimates for the period 2020 to 2023 are smaller, but the main conclusions of our original article still apply.

Our recent article [1] estimated excess mortality in Aotearoa New Zealand between 2020 and 2023 using official Statistics NZ estimates for the resident population based on the 2018 census [2], which was the best data available at the time the research was carried out (November 2024). Since our study was published, Statistics NZ have released updated population estimates for 30 June 2018 onwards based on the 2023 census and 2023 post-enumeration survey [3] (see Supplementary Material sec. 1 for details). This has enabled us to update our results using the most up-to-date population data [4].

Due to intercensal discrepancy, the new population estimates for 2020 to 2023 are lower than those used in our original study (Supplementary Figure S1), with the largest percentage decrease occurring in the over-90-years-old group (Supplementary Figure S2). As a result, expected deaths are lower and our updated excess mortality estimate is 1.3 percentage points higher than that reported in the original article (see Table 1), although the 95% confidence intervals (CIs) are overlapping and the new central estimate is within the original CIs [1]. Our updated estimate for excess mortality of 2.0% (95% CI [0.5%, 3.3%]) implies we can be confident that more deaths occurred than would have been expected in the absence of a pandemic. However, this estimate for excess mortality remains low in comparison to many peer countries, which experienced excess mortality rates close to 10% [5]. Updated versions of the Figures and Tables that appeared in the original article are provided in Supplementary Material. All the updated results used the same deaths data and the same computer code as in the original analysis. Data and code to reproduce these updated results are publicly available at [6]; the data and code used in the original analysis continue to be available at [7].

The new results do not qualitatively change any of the main conclusions of our original article: (i) excess mortality during the Covid-19 pandemic in Aotearoa New Zealand was very low by international standards; (ii) the timing and age-distribution of excess mortality was highly correlated with that of reported Covid-19 deaths (Supplementary Figures S5-S6); (iii) methods that fail to account for age will tend to overestimate excess mortality in Aotearoa New Zealand during this period. It should be noted that the method of Gibson [8], which ignores age, will also produce a larger estimated excess with the updated population estimates.

**Table 1.** Estimated excess mortality in New Zealand in the four-year period from 2020 to 2023 (mean and 95% confidence intervals). Excess mortality was calculated as actual deaths minus expected deaths estimated using a six-year pre-pandemic baseline (2014-2019) using the same quasi-Poisson regression model described in our original article [1]. The first column of results shows the number of excess deaths, the second shows the number of excess deaths as a percentage of the expected number of deaths. For results with different baselines, see Supplementary Tables S1-S2.

|  | Excess mortality (number of excess deaths) | Excess mortality (%) |
|---|---|---|
| Original estimate | 1040 [-1134, 2927] | 0.7% [-0.8%, 2.0%] |
| Updated estimate | 2849 [705, 4712] | 2.0% [0.5%, 3.3%] |

# Estimating excess mortality during the Covid-19 pandemic in Aotearoa New Zealand: Addendum

# Supplementary Material

Michael J. Plank, Pubudu Senanayake, Richard Lyon

1. **Summary of changes to the Statistics NZ estimated resident population**

Our model used quarterly age- and sex-specific estimated resident population (ERP) for the period 2010-Q1 to 2023-Q4 published by Statistics NZ. The ERP, at any given point in time, is an estimate based on a starting "base population" and accounting for subsequently observed births, deaths (natural increase/decrease) and net migration [1]. The base population is the coverage-adjusted resident population established from the most recent census, its associated post-enumeration survey (PES), and an estimate of residents overseas during the census [2].

The base population is updated after each census, in a process termed "rebasing" [1]. At the time of our original study, the most recent available population estimates were based on Census 2018 [3]. Estimates based on Census 2023 are now available [4] and this has resulted in revisions to the estimated population for 30 June 2018 onwards.

The total population estimate for 30 June 2023 was revised downwards by 44,500 (0.9%) compared to the previous estimate for 30 June 2023 based on Census 2018 [4]. This intercensal discrepancy arises as a result of inaccuracies in: census counts; adjustments to derive population estimates from census counts; and components of population change (births, deaths and migration). The total net intercensal discrepancy between 2018 and 2023 of -8,900 per year is comparable in magnitude to previous intercensal discrepancies [4]. However, unlike the 2013 to 2018 intercensal discrepancy, which was concentrated in the 15 to 60-year-old age bracket [3], the 2018 to 2023 intercensal discrepancy was concentrated in older age groups [4], and hence had a larger impact on expected deaths according to our model.

2. Supplementary Tables and Figures

| Baseline period | Excess (QPR) | Excess (SMR-LR) |
| --- | --- | --- |
| 2016-2019 | -316 [-3472, 2349] | -9 |
| 2015-2019 | 2118 [ -303, 4369] | 2325 |
| 2014-2019 | 2849 [705, 4712] | 3005 |
| 2013-2019 | 1168 [-711, 2889] | 1031 |
| 2012-2019 | 2532 [894, 4018] | 2661 |
| 2011-2019 | 4073 [2603, 5423] | 4375 |
| 2010-2019 | 3298 [1925, 4527] | 3430 |

**Supplementary Table S1**. Updated estimates of cumulative excess mortality for the 4-year period 2020 to 2023 with different length baselines. In each row of the table, the models were fitted to deaths data in a period of between 4 and 10 years ending in 2019. Results are shown for the quasi-Poisson regression (QPR) model (mean and 95% confidence interval) and the standardized mortality rate linear regression (SMR-LR) model.

| Baseline period | Excess (QPR) | Excess (SMR-LR) |
| --- | --- | --- |
| 2016-2019 | -1124 [-3199, 622] | -923 |
| 2015-2019 | 428 [-1198, 1937] | 577 |
| 2014-2019 | 899 [-556, 2153] | 1021 |
| 2013-2019 | -202 [-1498, 969] | -282 |
| 2012-2019 | 703 [-450, 1729] | 804 |
| 2011-2019 | 1732 [700, 2674] | 1957 |
| 2010-2019 | 1214 [254, 2070] | 1317 |

**Supplementary Table S2.** Updated estimates of cumulative excess mortality for the 3-year period 2020 to 2022 with different length baselines. In each row of the table, the models were fitted to deaths data in a period of between 4 and 10 years ending in 2019. Results are shown for the quasi-Poisson regression (QPR) model (mean and 95% confidence interval) and the standardized mortality rate linear regression (SMR-LR) model.

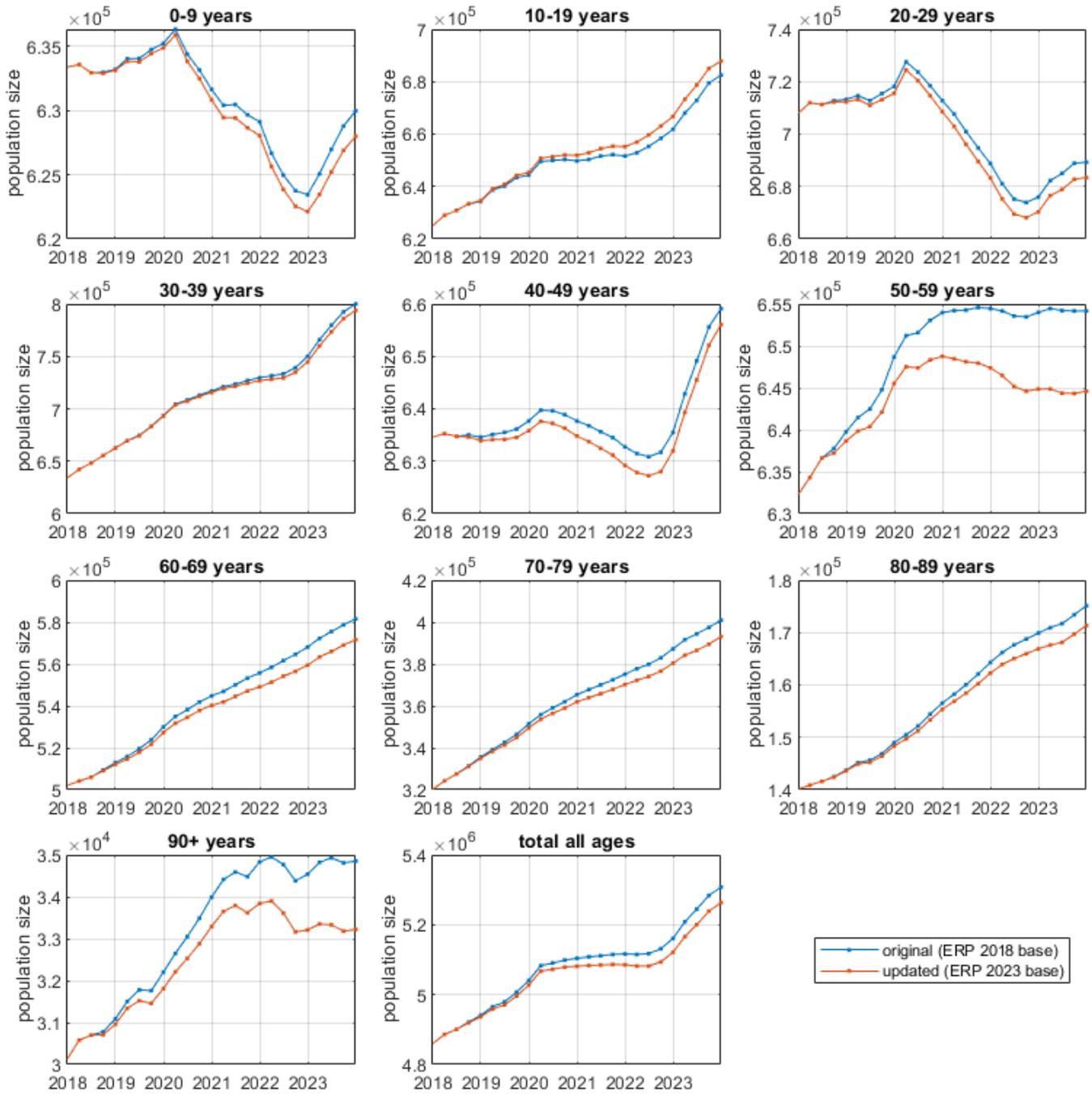

**Supplementary Figure S1.** Stats NZ estimated resident population (ERP) in ten-year age bands according to the data used in the original study (2018 census base, blue) and the updated data (2023 census base, red). Note the different vertical axis scales in the different plots.

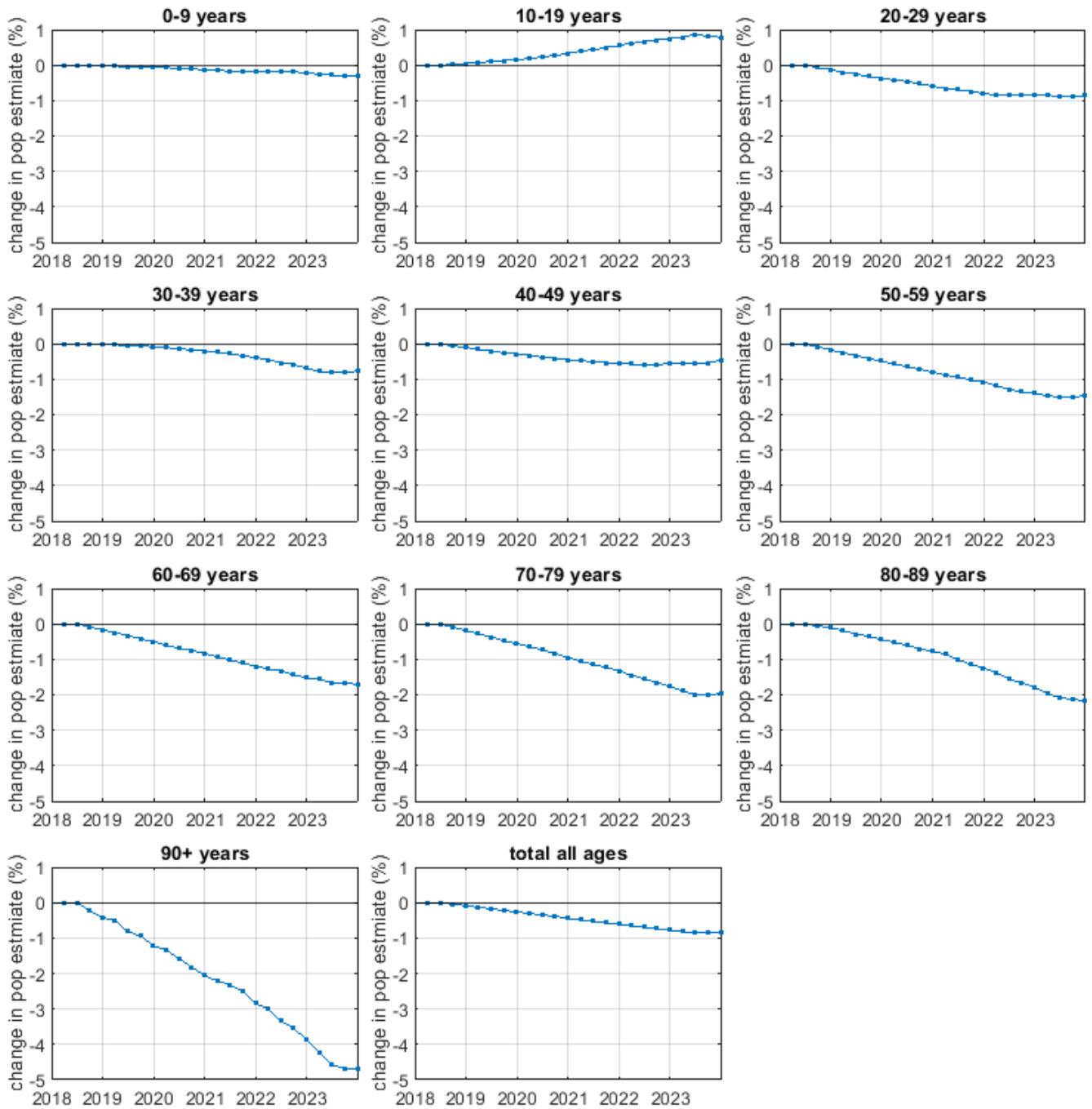

**Supplementary Figure S2.** Relative difference (%) between the 2013 base and 2018 base StatsNZ estimated resident population size in ten-year age bands, calculated as $100 \times \frac{(2023\ \text{estimate}) - (2018\ \text{estimate})}{(2018\ \text{estimate})}$. Negative values indicate that the updated population size estimate (2023 base) is smaller than the data used in the original study (2018 base).

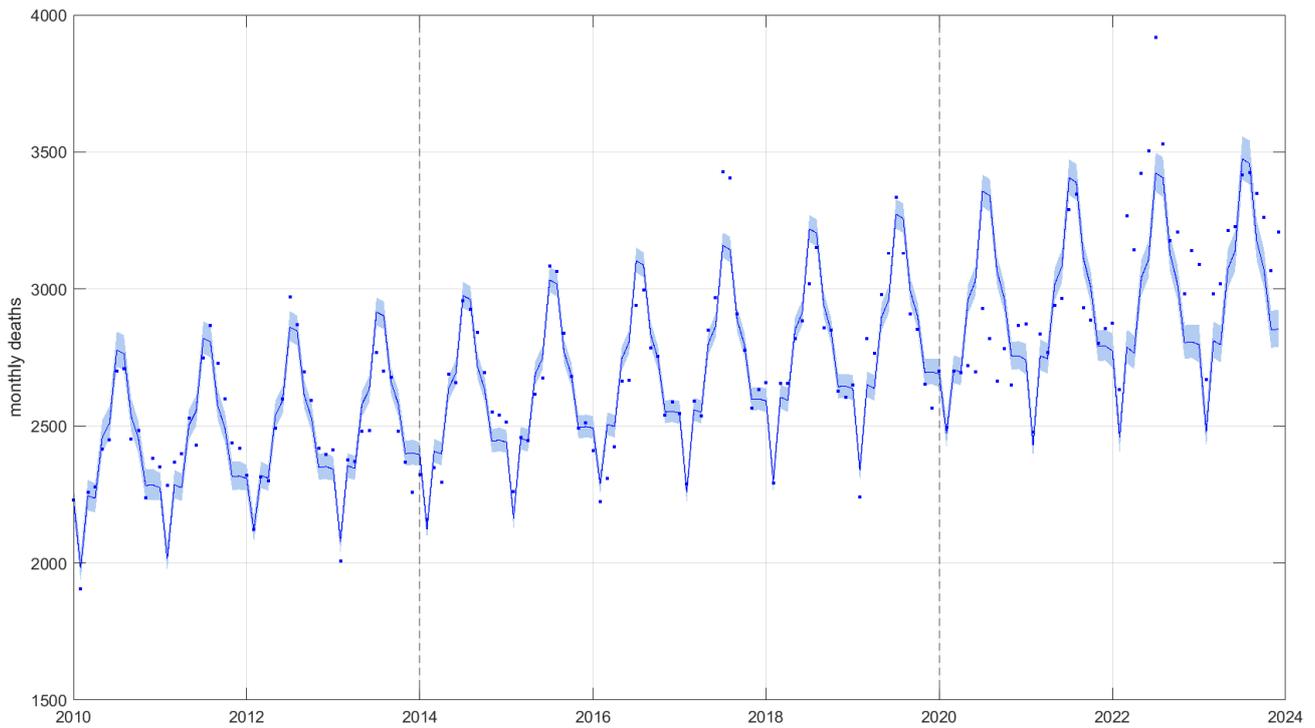

**Supplementary Figure S3.** Total monthly all-cause deaths in the New Zealand resident population from January 2010 to December 2023 (points) along with updated results from the quasi- Poisson regression model fitted to data on monthly deaths from January 2014 to December 2019. Vertical dashed lines show the fitting period. Solid curve shows the mean monthly deaths according to the fitted model; shaded band shows the 95% confidence interval (CI). Excess mortality is estimated as the difference between actual deaths for 2020–23 and expected deaths for 2020–23 according to the model (mean and 95% CI).

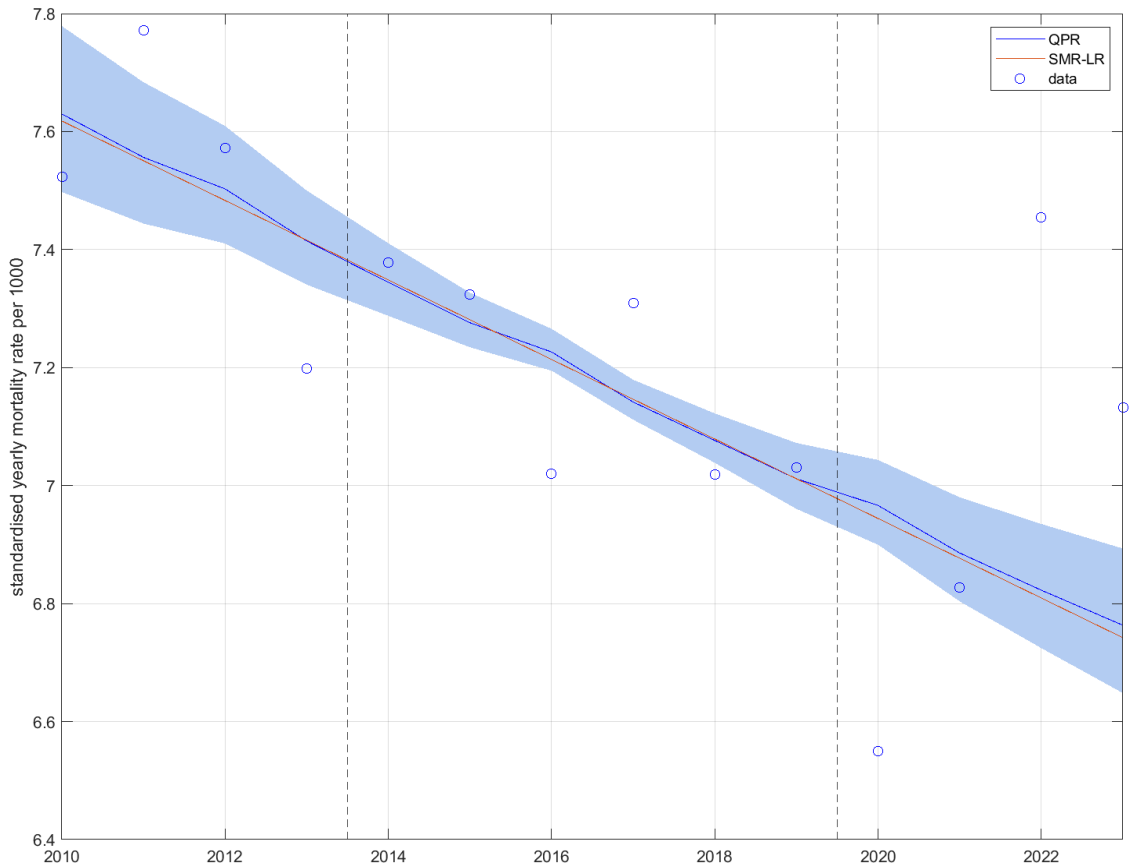

**Supplementary Figure S4.** Updated results for the age- and sex-standardized yearly all-cause mortality rate per 1000 people in the New Zealand resident population (open circles), with the results of the quasi-Poisson regression (QPR) model (solid blue curve = mean, shaded blue band = 95% confidence interval) and the standardized mortality rate linear regression (SMR-LR) model (solid red). Both models were fitted to data from 1 January 2014 to 31 December 2019 (indicated by the vertical dashed lines). All calculations use the first quarter of 2021 as the standard population.

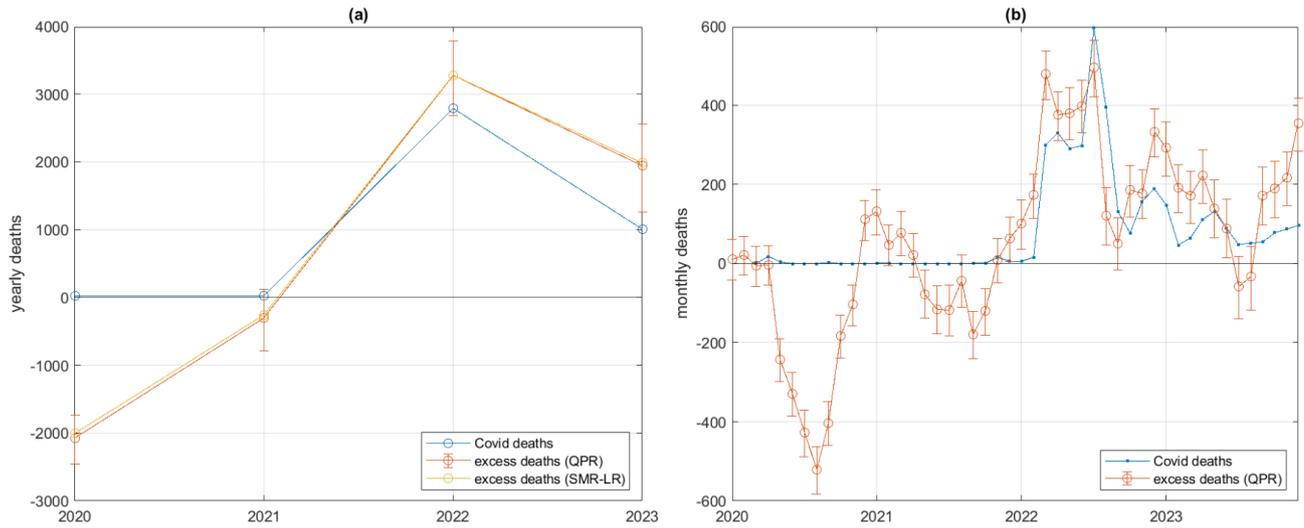

**Supplementary Figure S5.** (a) Yearly Covid-19-attributed deaths (blue) along with updated results for excess deaths according to the quasi-Poisson regression (QPR) model (red) and the standardized mortality rate linear regression (SMR-LR) model (yellow). (b) Monthly Covid-19-attributed deaths (blue points) along with updated results for excess deaths according to the QPR model (red open circles). Error bars show the 95% confidence interval for the QPR model.

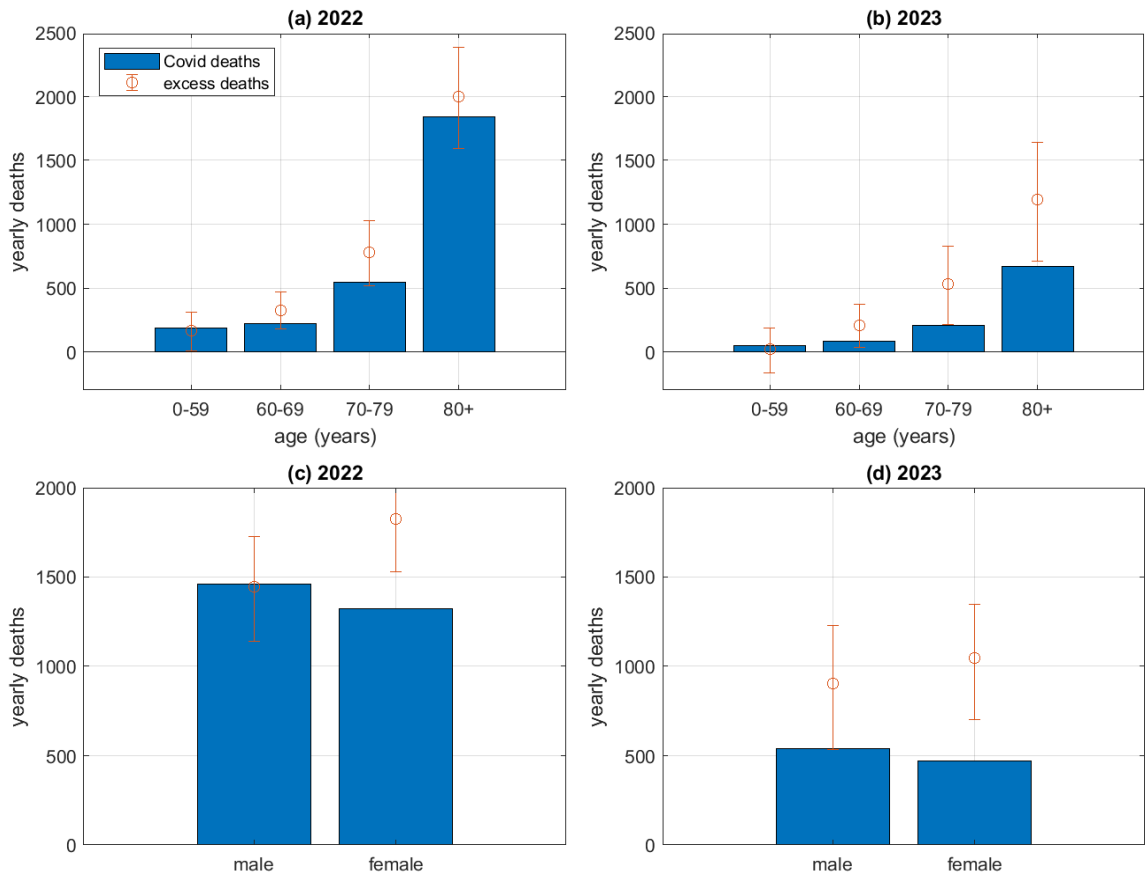

**Supplementary Figure S6.** Yearly Covid-19-attributed deaths (blue bars) and updated results for excess deaths (mean and 95% confidence interval) according to the quasi-Poisson model (open red circles and error bars) disaggregated by age (a, b) or sex (c, d) in 2022 and 2023.

**Supplementary references**